\documentstyle[11pt,epsfig]{article}
\title{\bf Generalized virial theorem in warped DGP brane-world}
\author{{Malihe Heydari-Fard \thanks{E-mail:
heydarifard@qom.ac.ir} \thanks{E-mail: m.heydarifard@sbu.ac.ir}
and Mohaddese Heydari-Fard}\thanks{E-mail: mheydarifard@qom.ac.ir}
\\{\small {{\emph{Department of Physics, The University of Qom, P.
O. Box 37155-1814, Qom, Iran}}}} }
\begin{document}
\maketitle 
\begin{abstract}
We generalize the virial theorem to the warped DGP brane-world
scenario and consider its implications on the virail mass. In this
theory the four-dimensional scalar curvature term is included in
the bulk action and the resulting four-dimensional effective
Einstein equation is augmented with extra terms which can be
interpreted as geometrical mass, contributing to the gravitational
energy. Estimating the geometrical mass ${\cal M}(r)$ using the
observational data, we show that these geometric terms may account
for the virial mass discrepancy in clusters of galaxies. Finally,
we obtain the radial velocity dispersion of galaxy clusters
$\sigma_{r}(r)$ and show that it is compatible with the radial
velocity dispersion profile of such clusters.
\vspace{5mm}
\vspace{0.5mm}
\end{abstract}

\section{Introduction}
In recent times theories with higher-dimensions have become
popular in high energy physics, specially in the context of
hierarchy problem and cosmology \cite{Witten}. In this scenario it
is purported that our four-dimensional universe is a subspace
called the brane, embedded in a higher-dimensional space-time
called the bulk. One of the most successful of such
higher-dimensional models is that proposed by Randall and Sundrum
whose the bulk has the geometry of an AdS space admitting $Z_2$
symmetry \cite{RS}. They were successful in explaining what is
known as the hierarchy problem: the enormous disparity between the
strength of the fundamental forces. The Randall Sundrum (RS)
scenario has had a great impact on our undersetting of the
universe and has brought higher-dimensional gravitational theories
to the fore. In certain RS type models, all matter and gauge
interactions reside on the brane while gravity can propagate into
the bulk. Using the Israel junction conditions \cite{Israel} and
the Gauss-Codazzi equations, one can obtain the field equations on
the brane, as employed by Shiromizu, Maeda and Sasaki (SMS)
\cite{SMS}. There are two very important results that arise from
the effective four-dimensional Einstein equations on the brane.
The first one is quadratic energy-momentum tensor, $\pi_{\mu\nu}$,
which is relevant in high energy and the second one is the
projected Weyl tensor, ${\cal E}_{\mu\nu}$, on the brane which is
responsible for carrying on the brane the contribution of the bulk
gravitational field. The cosmological evolution of such a brane
universe has been extensively investigated and effects such as a
quadratic density term in the Friedmann equations have been found
\cite{review1, Maartens, review2}.

An alternative scenario was subsequently proposed by Dvali,
Gabadadze and Porrati (DGP) \cite{Dvali}. The DGP proposal rests
on the key assumption of the presence of a four-dimensional Ricci
scalar in the bulk action. There are two main reasons that make
this model phenomenologically appealing. First, it predicts that
four-dimensional Newtonian gravity on a brane-world is regained at
distances shorter than a given crossover scale $r_c$ (high energy
limit), whereas five-dimensional effects become manifest above
that scale (low energy limit) \cite{Gabadadze}. Second, the model
can explain late-time acceleration without having to invoke a
cosmological constant or quintessential matter \cite{Deffayet}. An
extension of the DGP brane-world scenario have been constructed by
Maeda, Mizuno and Torii which is the combination of the RS II
model and DGP model \cite{maeda}. In this combination, an induced
curvature term appears on the brane in the RS II model. This model
has been called the warped DGP brane-world in literature
\cite{Cai}. In this paper, we consider the effective gravitational
field equations within the context of the warped DGP brane-world
model and obtain the spherically symmetric equations in this
scenario. So much for the success of the DGP model, a word of
caution is in order; the theory predicts the existence of
ghost-like excitations. Many scenarios have been undertaken to
explain away such ghosts, but as yet no satisfactory solution
exists. The interested reader should consult \cite{ghost} for
further insight. We do not discuss such excitations since our aim
lies in studying the virial mass discrepancy in warped DGP models.

Modern astrophysical and cosmological models are faced with two
severe theoretical difficulties which can be summarized as dark
energy and dark matter problems. The problem of dark matter is a
longstanding problem in modern astrophysics. Two important
observational issues, the behavior of the galactic rotation curves
and the mass discrepancy in clusters of galaxies led to the
necessity of considering the existence of dark matter at the
galactic and extra-galactic scales \cite{book}. The total mass of
a cluster can be estimated in two ways. One can apply the virial
theorem to estimate the total dynamic mass $M_{V}$ of a rich
galaxy cluster from measurements of the velocities of the member
galaxies and the cluster radius from the volume they occupy. The
second is obtained by separately estimating the mass of each
individual members and summing them up to give a total baryonic
mass $M$. It is always found that $M_V$ is greater than $M$. This
is known as the \emph{\textbf{missing mass problem}}.
Nevertheless, the existence of dark matter was not firmly
established until time when the measurement of the rotational
velocity of stars and gas orbiting at a distance $r$ from the
galactic center was performed. Observations show that the
rotational velocity increase near the center of galaxy and remain
nearly constant. This discrepancy between the observed rotation
velocity curves and the theoretical prediction from Newtonian
mechanics is known as \emph{\textbf{galactic rotation curves
problem}} (Figure 1). These discrepancies are explained by
postulating that every galaxy and cluster of galaxy is embedded in
a halo made up of some dark matter \cite{book}. To deal with the
question of dark matter, a great number of efforts has been
concentrated on various modifications to the Einstein and the
Newtonian gravity \cite{Mak1, Mak2, Mak3, Sefied, Shahidi}.
Several theoretical models, based on a modification of Newton's
law or of general relativity, have been proposed to explain the
behavior of the galactic rotation curves. In the modified
Newtonian dynamics (MOND) theory which has been proposed by
Milgrom \cite{Milgrom}, the Poisson equation for the gravitational
potential, $\nabla^2\phi = 4\pi G \rho$, is replaced by an
equation of the form $\nabla[\mu(x)(|\nabla\phi|/a_0)] = 4\pi G
\rho$, where $a_0$ is a fixed constant and $\mu(x)$ a function
satisfying the conditions $\mu(x) = x$ for $x << 1$ and $\mu(x) =
1$ for $x >> 1$. The force law, giving the acceleration $a$ of a
test particle becomes $a = a_N$ for $a_N >> a_0$ and $a =
\sqrt{a_Na_0}$ for $a_N << a_0$, where $a_N$ is the usual
Newtonian acceleration. The rotation curves of the galaxies are
predicted to be flat, and they can be calculated once the
distribution of the baryonic matter is known. A relativistic MOND
inspired theory was developed by Bekenstein \cite{Bekenstein}. In
this theory gravitation is mediated by a metric, a scalar field
and a four-vector field, all three dynamical. For alternative
theoretical models to explain the galactic rotation curves, see
\cite{Mannheim}. One other such modification is that of the RS
brane-world scenario \cite{Harko}. It has been argued that a
modified theory of gravity based on the RS brane-world scenario
can explain the observations of the galactic rotation curve of
spiral galaxies and the virial theorem mass discrepancy in
clusters of galaxies without introducing any additional hypothesis
\cite{Harko}.

\begin{figure}
\centerline{\begin{tabular}{ccc} \epsfig{figure=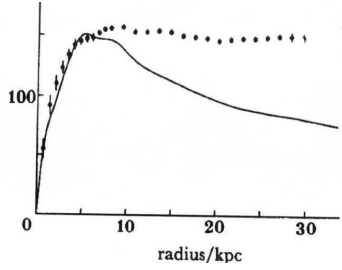,width=5cm}
\end{tabular} } \caption{\footnotesize Observed
rotation velocity curve of the NGC3198 (dotted line) and the
prediction from Newtonian theory (solid line).}\label{fig1}
\end{figure}

Our main purpose in this paper is to obtain the generalized form
of the virial theorem in Warped DGP brane-world model by using the
collision-less Boltzmann equation. In what follows, we first give
a brief review of warped DGP brane-world model and the
gravitational field equations are derived in this model. Next, we
use the relativistic Boltzmann equation to derive the virial
theorem which is modified by an extra term which may be used to
explain the virial mass discrepancy in clusters of galaxies.
Finally, we identify the geometrical mass of cluster in terms of
the observable quantities and obtain the radial velocity
dispersion of galaxy clusters.

\section{Effective field equations on warped DGP brane}
Let us start by presenting the model used in our calculation
\cite{maeda}. Consider a five-dimensional space-time with a
four-dimensional brane is located at $Y(X^A)=0$, where $X^A$,
($A=0,1,2,3,4$) are five-dimensional coordinates. The effective
action is given by
\begin{eqnarray}
{\cal S} = {\cal S}_{bulk}+{\cal S}_{brane},
\end{eqnarray}
where
\begin{eqnarray}
{\cal S}_{bulk} = \int d^{5}X\sqrt{-{\cal
G}}\left[\frac{1}{2\kappa_5^2}{\cal R}+{\cal
L}_{m}^{(5)}\right],\label{0}
\end{eqnarray}
and
\begin{eqnarray}
{\cal S}_{brane} = \int_{Y=0}d^{4}x
\sqrt{-g}\left[\frac{1}{\kappa_5^2}K^{\pm}+{\cal
L}_{brane}(g_{\alpha\beta},\psi)\right],\label{1}
\end{eqnarray}
where $\kappa_5^2=8\pi G_5$ is the five-dimensional gravitational
constant, ${\cal R}$ and ${\cal L}_{m}^{(5)}$ are the
five-dimensional scalar curvature and the matter Lagrangian in the
bulk, respectively. Also, $x^{\mu}$, ($\mu=0,1,2,3$) are the
induced four-dimensional coordinates on the brane, $K^{\pm}$ is
the trace of extrinsic curvature on either side of the brane
\cite{G,H} and ${\cal L}_{brane}(g_{\alpha\beta},\psi)$ is the
effective four-dimensional Lagrangian, which is given by a generic
functional of the brane metric $g_{\alpha\beta}$ and matter
fields.

The five-dimensional Einstein field equations are given by
\begin{eqnarray}
{\cal R}_{AB}-\frac{1}{2}{\cal R}{\cal G}_{AB} =
{\kappa_5^2}\left[T^{(5)}_{AB}+\delta(Y)\tau_{AB}\right],\label{2}
\end{eqnarray}
where
\begin{eqnarray}
T^{(5)}_{AB}\equiv-2\frac{\delta {\cal L}^{(5)}_{m}}{\delta{\cal
G}^{AB}}+{\cal G}_{AB}{\cal L}^{(5)}_{m},\label{3}
\end{eqnarray}
and
\begin{eqnarray}
\tau_{\mu\nu}\equiv-2\frac{\delta {\cal L}_{brane}}{\delta
g^{\mu\nu}}+g_{\mu\nu}{\cal L}_{brane}.\label{4}
\end{eqnarray}
We study the case with an induced gravity on the brane due to
quantum corrections \cite{Dvali}. The interaction between bulk
gravity and the matter on the brane induces gravity on the brane
through its quantum effects. If we take into account quantum
effects of matter fields confined on the brane, the gravitational
action on the brane is modified as
\begin{eqnarray}
{\cal L}_{brane}(g_{\alpha\beta},\psi) =
\frac{\mu^2}{2}R-\lambda_b+{\cal L}_{m},\label{5}
\end{eqnarray}
where $\mu$ is a mass scale which may correspond to the
four-dimensional Planck mass, $\lambda_b$ is the tension of the
brane and ${\cal L}_m$ presents the Lagrangian of the matter
fields on the brane. We note that for $\lambda_b=0$ and
$\Lambda^{(5)}=0$ action (1) gives the DGP model and gives the RS
II model if $\mu=0$.

We obtain the gravitational field equations on the brane-world as
\cite{SMS}
\begin{eqnarray}
G_{\mu\nu} = \frac{2\kappa_5^2}{3}\left[T_{AB}^{(5)}g^A_{\mu}
g^B_{\nu}+g_{\mu\nu}\left(T_{AB}^{(5)}n^An^B-\frac{1}{4}T^{(5)}\right)\right]+\kappa_{5}^2\pi_{\mu\nu}-{\cal
E}_{\mu\nu},\label{6}
\end{eqnarray}
\begin{eqnarray}
\nabla_{\nu}\tau_{\mu}^{\nu} =
-2T_{AB}^{(5)}n^{A}g^{B}_{\mu},\label{7}
\end{eqnarray}
where $\nabla_{\upsilon}$ is the covariant derivative with respect
to $g_{\mu\nu}$ and the quadratic correction has the form
\begin{eqnarray}
\pi_{\mu\nu} =
-\frac{1}{4}\tau_{\mu\alpha}\tau_{\nu}^{\alpha}+\frac{1}{12}\tau\tau_{\mu\nu}+
\frac{1}{8}g_{\mu\nu}\tau^{\alpha\beta}\tau_{\alpha\beta}-\frac{1}{24}g_{\mu\nu}\tau^2,\label{8}
\end{eqnarray}
and the projection of the bulk Weyl tensor to the surface
orthogonal to $n^A$ is given by
\begin{eqnarray}
{\cal E}_{\mu\nu} = C^{(5)}_{ABCD}n^An^Bg^C_{\,\,\,\mu}
g^D_{\,\,\,\nu}.\label{9}
\end{eqnarray}
The symmetry properties of ${\cal E}_{\mu\nu}$ imply that in
general we can decompose it irreducibly with respect to a chosen
4-velocity field $v^{\mu}$ as \cite{Maartens}
\begin{eqnarray}
{\cal E}_{\mu\nu} = -\left(\frac{\kappa_5}{\kappa_4}\right)^4
\left[U(v_{\mu}v_{\nu} + \frac{1}{3}h_{\mu\nu})+ 2
Q_{(\mu}v_{\nu)} + P_{\mu\nu}\right],
\end{eqnarray}
where $h_{\mu\nu} = g_{\mu\nu}+v_{\mu}v_{\nu}$ projects orthogonal
to $v_{\mu}$ and the factor $\left({\kappa_5}/{\kappa_4}\right)$
is introduced for dimension reasons. Here $$U =
-\left(\frac{\kappa_4}{\kappa_5}\right)^4{\cal
E}_{\mu\nu}v^{\mu}v^{\nu},$$ is an effective non-local energy
density or ``dark radiation'' term on the brane, arising from the
free gravitational field in the bulk, $Q_{\mu} =
\left({\kappa_4}/{\kappa_5}\right)^4h^{\alpha}_{\mu}{\cal
E}_{\alpha\beta}v^{\beta}$ is an effective non-local energy flux
and $$P_{\mu\nu} =
-\left(\frac{\kappa_4}{\kappa_5}\right)^4\left[h_{(\mu}^{\,\,\,\,\alpha}h_{\nu)}^{\,\,\,\,\beta}-
\frac{1}{3}h_{\mu\nu}h^{\alpha\beta}\right]{\cal
E}_{\alpha\beta},$$ is a spatial, symmetric and trace-free tensor.
In what follows for the static spherically symmetric brane we have
$Q_{\mu} = 0$, and we may choose $P_{\mu\nu} =
P(r)(r_{\mu}r_{\nu}-\frac{1}{3}h_{\mu\nu})$, where the ``dark
pressure'' $P(r)$ is a scalar functions of the radial distance
$r$, $r_{\mu}$ is a unit radial vector and at any point on the
brane in inertial frame $v^{\mu} = \delta^{\mu}_{0}, h_{\mu\nu} =
diag(0, 1, 1, 1)$ \cite{Dadhich}.

In order to find the basic field equations on the brane with
induced gravity, we have to obtain the energy-momentum tensor of
the brane $\tau_{\mu\nu}$, given by definition (\ref{4}) from the
Lagrangian (\ref{5}), yielding
\begin{eqnarray}
\tau_{\nu}^{\mu} =
-\lambda_b\delta^{\mu}_{\nu}+T^{\mu}_{\nu}-\mu^2G_{\nu}^{\mu}.\label{10}
\end{eqnarray}
Assuming that the five-dimensional bulk space includes only a
cosmological constant $\Lambda^{(5)}$ and inserting equation
(\ref{10}) into equation (\ref{6}), we find the effective field
equations for four-dimensional metric $g_{\mu\nu}$ as
\begin{eqnarray}
\left(1+\frac{\lambda_b}{6}\kappa_{5}^4\mu^2\right)G_{\mu\nu}=
\frac{1}{6}\lambda_b \kappa_{5}^{4}T_{\mu\nu}-\Lambda_4
g_{\mu\nu}-\kappa_{5}^{4}\mu^2{
K}_{\mu\nu\alpha\beta}G^{\alpha\beta}+{\kappa}_{5}^{4}\left[\pi_{\mu\nu}^{(T)}+\mu^4\pi_{\mu\nu}^{(G)}\right]-{\cal
E}_{\mu\nu},\label{11}
\end{eqnarray}
where
\begin{eqnarray}
K_{\mu\nu\rho\sigma} =
\frac{1}{4}\left(g_{\mu\nu}T_{\rho\sigma}-g_{\mu\rho}T_{\nu\sigma}-g_{\nu\sigma}T_{\mu\rho}\right)+\frac{1}{12}
\left[T_{\mu\nu}g_{\rho\sigma}+T(g_{\mu\rho}g_{\nu\sigma}-g_{\mu\nu}g_{\rho\sigma})\right],\label{12}
\end{eqnarray}
\begin{eqnarray}
\pi^{(T)}_{\mu\nu} =
-\frac{1}{4}T_{\mu\alpha}T^{\alpha}_{\nu}+\frac{1}{12}TT_{\mu\nu}
+\frac{1}{8}g_{\mu\nu}T_{\alpha\beta}T^{\alpha\beta}-\frac{1}{24}g_{\mu\nu}T^2,\label{13}
\end{eqnarray}
\begin{eqnarray}
\pi^{(G)}_{\mu\nu} =
-\frac{1}{4}G_{\mu\alpha}G^{\alpha}_{\nu}+\frac{1}{12}GG_{\mu\nu}
+\frac{1}{8}g_{\mu\nu}G_{\alpha\beta}G^{\alpha\beta}-\frac{1}{24}g_{\mu\nu}G^2,\label{14}
\end{eqnarray}
and the effective cosmological constant on the brane is given by
\begin{eqnarray}
\Lambda_4 =
\frac{\kappa_5^2}{2}\left[\Lambda^{(5)}+\frac{1}{6}\kappa_5^2\lambda_b^2\right].\label{15}
\end{eqnarray}
We note that for $\mu=0$ these equations are exactly the same
effective equations as in the reference \cite{SMS}.

\section{Field equations for a cluster including identical and
collision-less point particles}

Now we consider an isolated and spherically symmetric cluster
being described by a static and spherically symmetric metric
\begin{eqnarray}
ds^2 =
-e^{\lambda(r)}dt^2+e^{\nu(r)}dr^2+r^2(d\theta^2+sin^2\theta
d\varphi^2).\label{16}
\end{eqnarray}
Suppose that the clusters are constructed from identical and
collision-less point particles (galaxies). This multi-particle
system can be described by a continuous non-negative function
$f_B(x^{\mu},p^{\mu})$, distribution function, which is defined
over the phase space. In terms of the distribution function the
energy-momentum tensor can be written as \cite{tetrad,Tetrad}
\begin{eqnarray}
T_{\mu\nu} = \int f_B m v_{\mu}v_{\nu} dv,\label{b10}
\end{eqnarray}
where $m$ is the mass of each galaxy, $v_{\mu}$ is the
four-velocity of the galaxy and $dv = \frac{1}{v_t}dv_r dv_\theta
dv_\varphi$ is the invariant volume element of the velocity space.
We also assume that the matter content of the bulk is just a
cosmological constant $\Lambda^{(5)}$ and the energy-momentum
tensor of the matter in a cluster of galaxies can be represented
in terms of spherically symmetric perfect fluid as
\begin{eqnarray}
T_{\mu\nu} = \left(p_b+\rho_b\right)v_\mu
v_\nu+p_bg_{\mu\nu},\label{B10}
\end{eqnarray}
where $v_\mu v^\mu = -1$. Use of equations (\ref{b10}) and
(\ref{B10}) leads to the following relation for $\rho_b$ and $p_b$
\begin{eqnarray}
\rho_{b} = \rho\langle{v_{t}^{2}}\rangle,\hspace{.5 cm}p_{b} =
\rho\langle{v_{r}^{2}}\rangle = \rho\langle{v_{\theta}^{2}}\rangle
= \rho\langle{v_\varphi^{2}}\rangle,\label{b11}
\end{eqnarray}
here $\langle{v_{t}^{2}}\rangle$ represents the usual macroscopic
averaging which is defined as $\langle{v_{t}^{2}}\rangle =
\frac{1}{\rho}\int f_B {v_{t}^{2}} m dv$ where $\rho$ is the mass
density and in term of $f_B$ is given by $\rho = \int f_B m dv $
\cite{Jackson, Dark matter}.

Using equation (\ref{11}), the gravitational field equations on
the brane become
\begin{eqnarray}
(1+\frac{\lambda_b}{6}\kappa_{5}^4\mu^2)\frac{e^{-\nu}}{r^2}\left(-1+r\nu^{'}+e^{\nu}\right)
=
\frac{1}{6}\lambda_b\kappa_{5}^{4}\rho_b+\Lambda_4+\kappa_{5}^{4}\mu^2{\cal
K}_0^{0}+\frac{\kappa_{5}^4}{12}\rho_b^2-\kappa_{5}^4\mu^4\pi_0^{0(G)}+\frac{6}{\kappa_4^2\lambda_b}U(r)
,\label{19}
\end{eqnarray}
\begin{eqnarray}
(1&+&\frac{\lambda_b}{6}\kappa_{5}^4\mu^2)\frac{e^{-\nu}}{r^2}\left(1+r\lambda^{'}-e^{\nu}\right)
=
\frac{1}{6}\lambda_b\kappa_{5}^{4}p_b-\Lambda_4-\kappa_{5}^{4}\mu^2{\cal
K}_1^{1}+\frac{\kappa_{5}^4}{12}\left(\rho_b^2+2
\rho_b p_b\right)+\kappa_{5}^4\mu^4\pi_1^{1(G)}\nonumber\\
&+&\frac{2}{\kappa_4^2\lambda_b}\left[U(r)+2P(r)\right]
,\label{20}
\end{eqnarray}
\begin{eqnarray}
(1&+&\frac{\lambda_b}{6}\kappa_{5}^4\mu^2)\frac{e^{-\nu}}{2r}\left(2\lambda^{'}-2\nu^{'}-\lambda^{'}\nu^{'}
r+2\lambda^{''}r+\lambda^{'2}r\right) =
\frac{1}{3}\lambda_b\kappa_{5}^{4}p_b-2\Lambda_4-2\kappa_{5}^{4}\mu^2{
K}_2^{2}+\frac{\kappa_{5}^4}{6}\left(\rho_b^2+2\rho_b
p_b\right)\nonumber\\
&+&2\kappa_{5}^4\mu^4\pi_2^{2(G)}
+\frac{4}{\kappa_4^2\lambda_b}[U(r)-P(r)] ,\label{21}
\end{eqnarray}
where $U(r)$ and $P(r)$ are the dark radiation and dark pressure
and ${K}_{\mu\nu\alpha\beta}G^{\alpha\beta} \equiv {\cal
K}_{\mu\nu}$ term is given by
\begin{eqnarray}\label{}
{\cal K}_0^{0} &=&
-\frac{e^{-\nu}}{8}\left[\frac{2\nu'}{r}\left(\rho_b+p_b\right)-4p_b\frac{\lambda'}{r}-\frac{2}{r^2}\left(\rho_b+p_b\right)
+p_b\lambda'\nu'-2p_b\lambda''-p_b\lambda'^2+\frac{2}{r^2}(\rho_b+p_b)e^{\nu}\right]\nonumber\\
&-&\frac{e^{-\nu}\left(\rho_b-3p_b\right)}{24}\left[\frac{2e^{\nu}}{r^2}+\frac{2\nu'}{r}-\frac{2}{r^2}-\frac{4\lambda'}{r}+\lambda'\nu'
-2\lambda''-\lambda'^2\right]\nonumber\\
&+&\frac{\rho_be^{-\nu}}{24}\left[\frac{4\nu'}{r}-\frac{4\lambda'}{r}
-\frac{4}{r^2}+\lambda'\nu'-2\lambda''-\lambda'^2+\frac{4e^{\nu}}{r^2}\right],
\end{eqnarray}
\begin{eqnarray}\label{}
{\cal K}_1^{1} &=&
-\frac{e^{-\nu}}{8}\left[p\lambda'\nu'-2p\lambda''-p\lambda'^2-\frac{2\nu'}{r}\left(\rho_b-p_b\right)+\frac{2}{r^2}\left(\rho_b+p_b\right)e^{\nu}
+\frac{2}{r^2}\left(\rho_b+p_b\right)\right]\nonumber\\
&-&\frac{e^{-\nu}\left(\rho_b-3p_b\right)}{24}\left[\frac{4\nu'}{r}+\frac{2e^{\nu}}{r^2}
-\frac{2}{r^2}-\frac{2\lambda'}{r}+\lambda'\nu'-2\lambda''-\lambda'^2\right]\nonumber\\
&-&\frac{p_be^{-\nu}}{24}\left[\frac{4\nu'}{r}+\frac{4e^\nu}{r^2}-\frac{4}{r^2}-
\frac{4\lambda'}{r}+\lambda'\nu'-2\lambda''-\lambda'^2\right],
\end{eqnarray}
\begin{eqnarray}\label{}
{\cal K}_2^{2} &=& {\cal K}_3^{3} =
\frac{e^{-\nu}}{4}\left[\frac{p_b\lambda'}{r}+\frac{\rho_b\nu'}{r}+\frac{\left(\rho_b-p_b\right)e^\nu}{r^2}-
\frac{\left(\rho_b-p_b\right)}{r^2}\right]
\nonumber\\
&-&\frac{e^{-\nu}\left(\rho_b-3p_b\right)}{48}\left[\frac{6\nu'}{r}+\frac{8e^\nu}{r^2}-\frac{8}{r^2}-
\frac{6\lambda'}{r}+\lambda'\nu'-2\lambda''-\lambda'^2\right]\nonumber\\
&-&\frac{p_be^{-\nu}}{24}\left[\frac{4\nu'}{r}+\frac{4e^\nu}{r^2}-\frac{4}{r^2}-\frac{4\lambda'}{r}+
\lambda'\nu'-2\lambda''-\lambda'^2\right],
\end{eqnarray}
and use of equation (\ref{14}) leads to the components of
$\pi_{\mu}^{\nu(G)}$ as
\begin{eqnarray}\label{22}
\pi_0^{0(G)} &=&
\frac{e^{-\nu}}{24}\left(\frac{-6\nu^{'}}{r^3}-\frac{2\lambda^{'}}{r^3}-\frac{\lambda^{'}\nu^{'}}{r^2}
+\frac{2\lambda^{''}}{r^2}+\frac{\lambda^{'
2}}{r^2}\right)+\frac{{e^{-2\nu}}}{192}(\frac{48\nu^{'}}{r^3}+\frac{16\lambda^{'}}{r^3}-
\frac{8\nu^{'}\lambda^{''}}{r}+\frac{16\lambda^{'}\nu^{'}}{r^2}\nonumber\\
&-&
{4\lambda^{'}\lambda^{''}\nu^{'}}-\frac{8\lambda^{'}\lambda^{''}}{r}+\frac{4\lambda^{'}\nu^{'2}}{r}+{
\lambda^{'2}\nu^{'2}}-{2\lambda^{'3}\nu^{'}}+{4\lambda^{''}\lambda^{'2}}-\frac{12\nu^{'2}}{r^2}
-\frac{4\lambda^{'3}}{r}\nonumber\\
&-&\frac{16\lambda^{''}}{r^2}+{4\lambda^{''2}}
+{\lambda^{'4}}-\frac{4\lambda^{'2}}{r^2}),
\end{eqnarray}
\begin{eqnarray}\label{23}
\pi_{1}^{1(G)} &=&
\frac{e^{-\nu}}{24}\left(\frac{2\nu^{'}}{r^3}+\frac{6\lambda^{'}}{r^3}-\frac{\lambda^{'}\nu^{'}}{r^2}
+\frac{2\lambda^{''}}{r^2}+\frac{\lambda^{'2}}{r^2}\right)+
\frac{e^{-2\nu}}{192}(-\frac{16\nu^{'}}{r^3}-\frac{48\lambda^{'}}{r^3}+
\frac{8\nu^{'}\lambda^{''}}{r}+\frac{16\lambda^{'}\nu^{'}}{r^2}\nonumber\\
&-&
{4\lambda^{'}\nu^{'}\lambda^{''}}+\frac{8\lambda^{'}\lambda^{''}}{r}-
\frac{4{\lambda^{'}\nu^{'2}}}{r}+{
\mu^{'2}\nu^{'2}}-{2\lambda^{'3}\nu^{'}}+{4\lambda^{''}\lambda^{'2}}
+\frac{4\nu^{'2}}{r^2}+\frac{4\lambda^{'3}}{r}\nonumber\\
&-&\frac{16\lambda^{''}}{r^2}+{4\lambda^{''2}}+{\lambda^{'4}}-\frac{20\lambda^{'2}}{r^2})
,
\end{eqnarray}
\begin{eqnarray}\label{24}
\pi_2^{2(G)} &=& \pi_3^{3(G)} =
-\frac{e^{-\nu}}{24}\left(\frac{4}{r^4}+\frac{\lambda^{'}\nu^{'}}{r^2}-\frac{2\lambda^{''}}{r^2}-\frac{\lambda^{'2}}{r^2}
-\frac{2e^{\nu}}{r^4}\right)-
\frac{{e^{-2\nu}}}{48}(-\frac{2\lambda^{'2}\nu^{'}}{r}-\frac{2\lambda^{''}\nu^{'}}{r}-\frac{4}{r^4}\nonumber\\
&-&\frac{10\lambda^{'}\nu^{'}}{r^2}+\frac{2\lambda^{'}\lambda^{''}}{r}+\frac{\lambda^{'}\nu^{'2}}{r}-\frac{2\nu^{'2}}{r^2}
+\frac{\lambda^{'3}}{r}+\frac{4\lambda^{''}}{r^2}),
\end{eqnarray}
where a prime represents differentiation with respect to $r$. In
the next section, we will investigate the influence of the bulk
effects on the dynamics of the galaxies in warped DGP brane-world
model.

\section{The virial theorem in warped DGP brane}
In order to derive the virial theorem for galaxy clusters, we have
to first write down the general relativistic Boltzmann equation
governing the evolution of the distribution function $f_B$. The
galaxies, which are treated as identical and collision-less point
particles, are described by this distribution function. For the
static spherically symmetric metric given by equation (\ref{16})
we introduce the following frame of orthonormal vectors
\cite{tetrad, Tetrad, Jackson}
\begin{eqnarray}
e^{(0)}_{\rho} = e^{\lambda/2}\delta^{0}_{\rho},\hspace{.5
cm}e^{(1)}_{\rho} = e^{\nu/2}\delta^{1}_{\rho},\hspace{.5
cm}e^{(2)}_{\rho} = r\delta^{2}_{\rho},\hspace{.5
cm}e^{(3)}_{\rho} = r\sin\theta\delta^{3}_{\rho},\label{b1}
\end{eqnarray}
where $g^{\mu\nu}e_{\mu}^{(a)}e_{\nu}^{(b)}=\eta^{(a)(b)}$. The
four-velocity $v^{\mu}$ of a typical galaxy with
$v^{\mu}v_{\mu}=-1$, in tetrad components is written as
\begin{eqnarray}
v^{(a)} = v^{\mu}e^{(a)}_{\mu},\hspace{.5 cm} a =
0,1,2,3.\label{b2}
\end{eqnarray}
The relativistic Boltzmann equation in tetrad components is given
by
\begin{eqnarray}
v^{(a)}e^{\rho}_{(a)}\frac{\partial f_B}{\partial
x^{\rho}}+\gamma^{(a)}_{(b)(c)}v^{(b)}v^{(c)}\frac{\partial
f_B}{\partial v^{(a)}} = 0,\label{b3}
\end{eqnarray}
where $f_B = f_B(x^{\mu},v^{(a)})$ and
$\gamma^{(a)}_{(b)(c)}=e^{(a)}_{\rho;\sigma}e^{\rho}_{(b)}e^{\sigma}_{(c)}$
are the distribution function and the Ricci rotation coefficients,
respectively. Assuming that the distribution function is only a
function of $r$, the relativistic Boltzmann equation becomes
\begin{eqnarray}
v_{r}\frac{\partial f_B}{\partial
r}-\left[\frac{v_{t}^{2}}{2}\frac{\partial \lambda}{\partial
r}-\frac{(v_{\theta}^{2}+v_{\varphi}^{2})}{r}\right]\frac{\partial
f_B}{\partial v_{r}}-\frac{v_{r}}{r}\left[v_{\theta}\frac{\partial
f_B}{\partial v_{\theta}}+v_{\varphi}\frac{\partial f_B}{\partial
v_{\varphi}}\right]-\frac{e^{\nu/2}v_{\varphi}}{r}\cot{\theta}\left[v_{\theta}\frac{\partial
f_B}{\partial v_{\varphi}}-v_{\varphi}\frac{\partial f_B}{\partial
v_{\theta}}\right] = 0,\label{b4}
\end{eqnarray}
where we have defined
\begin{equation}\label{}
v^{(0)} = v_t,\hspace{.5 cm}v^{(1)} = v_r,\hspace{.5 cm}v^{(2)} =
v_\theta,\hspace{.5 cm}v^{(3)} = v_\varphi.\label{b5}
\end{equation}
Since we have assumed the system to be spherically symmetric, the
term proportional to cot$\theta$ must be zero. Multiplying
equation (\ref{b4}) by $m v_r dv$ where $dv = \frac{1}{v_t}dv_r
dv_\theta dv_\varphi$, and integrating over the velocity space and
assuming that the distribution function vanishes rapidly as the
velocities tend to $\pm \infty$, we obtain
\begin{eqnarray}
r\frac{\partial}{\partial r}\left[\rho\langle { v_r^2 }
\rangle\right]+\frac{1}{2}\rho\left[\langle{v_t^2}\rangle+\langle{v_r^2}\rangle\right]
r\frac{\partial \lambda}{\partial
r}-\rho\left[\langle{v_\theta^2}\rangle+\langle{v_\varphi^2}\rangle-2\langle{v_r^2}\rangle\right]
 = 0,\label{b6}
\end{eqnarray}
where $\rho$ is the mass density and $\langle v_r^2 \rangle$
represents the average value of $v_r^2$. Multiplying equation
(\ref{b6}) by $4\pi r^2$ and integrating over the cluster leads to
\begin{eqnarray}
&&\int^{R}_{0}4\pi\rho\left[\langle{v_r^2}\rangle
+\langle{v_\theta^2}\rangle+\langle{v_\varphi^2}\rangle\right]r^2
dr-\frac{1}{2}\int^{R}_{0}4\pi r^3\rho
\left[\langle{v_t^2}\rangle+\langle{v_r^2}\rangle\right]\frac{\partial
\lambda}{\partial r}dr = 0.\label{b7}
\end{eqnarray}
This equation is reduced to
\begin{eqnarray}
2K-\frac{1}{2}\int^{R}_{0}4\pi
r^3\rho\left[\langle{v_t^2}\rangle+\langle{v_r^2}\rangle\right]\frac{\partial
\lambda}{\partial r}dr = 0,\label{b8}
\end{eqnarray}
where the total kinetic energy of the galaxies is defined as
\begin{eqnarray}
K = \int^{R}_{0}2\pi\rho\left[\langle{v_r^2}\rangle+
\langle{v_\theta^2}\rangle+\langle{v_\varphi^2}\rangle\right]r^2
dr.\label{b9}
\end{eqnarray}
Now, using these relations and adding the gravitational field
equations (\ref{19})-(\ref{21}) and equation (\ref{b11}) we find

\begin{eqnarray}
&&\left(1+\frac{\lambda_b}{6}\kappa_{5}^4\mu^2\right)e^{-\nu}\left(\frac{\lambda^{'}}{r}-\frac{\lambda^{'}\nu^{'}}{4}
+\frac{\lambda^{''}}{2}+ \frac{\lambda^{'2}}{4}\right) =
\frac{\kappa_4^2}{2}\rho\left[\langle{v_t^2}\rangle+\langle{v_r^2}\rangle+
\langle{v_\theta^2}\rangle+\langle{v_\varphi^2}\rangle\right]\nonumber\\
&+&\frac{\kappa_4^2}{2\lambda_b}\rho^2\left[\langle{v_t^2}\rangle^2+\langle{v_r^2}\rangle^2+
\langle{v_\theta^2}\rangle^2+\langle{v_\varphi^2}\rangle^2\right]-\Lambda_4+\frac{\kappa_5^4\mu^2}{2}[{\cal
K}_0^{0}-{\cal K}_1^{1}-2{\cal
K}_2^{2}]\nonumber\\&-&\frac{\kappa_5^4\mu^4}{2}[\pi_0^{0(G)}
-\pi_1^{1(G)}-2\pi_2^{2(G)}]+\frac{6}{\kappa_4^2\lambda_b}U(r).\label{b122}
\end{eqnarray}
where $\frac{1}{6}\lambda_b\kappa_{5}^{4} = \kappa_4^2$. In order
to obtain the generalized virial theorem we have to use some
approximations. First, since the dispersion of the velocity of
galaxies in the clusters is of the order $600-1000 $$km / s$,
i.e., $(\frac{v}{c})^2\approx 4\times10^{-6} - 1.11\times10^{-5}
\ll 1$, we can neglect the relativistic effects in the
relativistic Boltzmann equation and use the small velocity limit
approximation. In other words, $\langle{v_r^2}\rangle\approx
\langle{v_\theta^2}\rangle\approx
\langle{v_\varphi^2}\rangle\ll\langle{v_t^2}\rangle\approx1$.
Second, the intensity of the gravitational effects can be
estimated from the ratio $G M/R$, which for typical clusters is of
the order of $10^{-6}\ll 1$. Therefore inside the galactic
clusters the gravitational field is weak and we can  use the weak
gravitational field approximation. Then the term proportional to
$\lambda'\nu'$ and ${\lambda'}^{2}$ in equation (\ref{b122}) may
be ignored. Thus, assuming that $e^\lambda\approx e^\nu\approx 1$
inside the cluster \cite{Harko}, we can write equations
(\ref{b122}) as
\begin{eqnarray}
(1&+&\frac{\lambda_b}{6}\kappa_{5}^4\mu^2)\left(\frac{\lambda^{'}}{r}
+\frac{\lambda^{''}}{2}\right) =
\frac{\kappa_4^2}{2}\rho\left[\langle{v_t^2}\rangle+\langle{v_r^2}\rangle+
\langle{v_\theta^2}\rangle+\langle{v_\varphi^2}\rangle\right]
+\frac{\kappa_4^2}{2\lambda_b}\rho^2\left[\langle{v_t^2}\rangle^2+\langle{v_r^2}\rangle^2+
\langle{v_\theta^2}\rangle^2+\langle{v_\varphi^2}\rangle^2\right]
\nonumber\\
&-&\Lambda_4-\frac{\kappa_5^4\mu^2}{6r}\rho\left[3\nu'\langle{v_t^2}\rangle+2\lambda'\langle{v_t^2}\rangle
+\lambda''
r\langle{v_t^2}\rangle+3\nu'\langle{v_r^2}\rangle\right]+\frac{6}{\kappa_4^2\lambda_b}U(r).\label{b12}
\end{eqnarray}
On the other hand, for clusters of galaxies the ratio of the
matter density and of the brane tension is much smaller than 1,
$\rho/\lambda_b << 1$, so that one can neglect the quadratic term
in the matter density in above equation. These conditions
certainly apply to test particles in stable circular motion around
galaxies, and to the galactic clusters. Thus, we can rewrite
equation (\ref{b12}) as
\begin{eqnarray}
\left(1+\frac{\lambda_b}{6}\kappa_{5}^4\mu^2\right)\frac{1}{2r^2}\frac{\partial}{\partial
r}\left(r^2\frac{\partial \lambda}{\partial r}\right) =
\frac{\kappa_4^2}{2}\rho-\Lambda_4+{\kappa_5^4\mu^2}[{\cal
P}(r)+3{\cal U}(r)]+\frac{6}{\kappa_4^2\lambda_b}U(r),\label{b13}
\end{eqnarray}
where
$${\cal U}(r) = -\frac{\rho\nu'}{6r},$$
$${\cal P}(r) = -\frac{\rho}{6r^2}\frac{\partial}{\partial
r}\left(r^2\frac{\partial \lambda}{\partial r}\right).
$$
Multiplying equation (\ref{b13}) by $r^2$ and integrating from 0
to $r$ yields
\begin{eqnarray}
\left(1+\frac{\lambda_b}{6}\kappa_{5}^4\mu^2\right)\frac{1}{2}\left(r^2\frac{\partial
\lambda}{\partial
r}\right)-\frac{\kappa_4^2}{8\pi}M(r)+\frac{1}{3}\Lambda_4r^3-\frac{\kappa_4^2}{8\pi}{
M}_{DGP}(r)-\frac{\kappa_4^2}{8\pi}{M}_{RS}(r) = 0.\label{b14}
\end{eqnarray}
The total baryonic mass and the geometrical masses of the system
are given by
\begin{eqnarray}
M(r) = 4\pi \int^r_0\rho(r') r^{'2}dr^{'},\label{b15}
\end{eqnarray}
and
\begin{eqnarray}
\kappa_4^2{M}_{RS}(r) = \frac{48\pi}{\kappa_4^2\lambda_b} \int^r_0
{U}(r') r^{'2}dr^{'},\label{M1}
\end{eqnarray}
\begin{eqnarray}
\kappa_4^2{M}_{DGP}(r) = 8\pi\kappa_5^4\mu^2 \int^r_0 [{\cal
P}(r')+3{\cal U}(r')] r^{'2}dr^{'},\label{M2}
\end{eqnarray}
where
\begin{equation}\label{M3}
{\cal M}(r) = M_{RS}(r)+M_{DGP}(r).
\end{equation}
Multiplying equation (\ref{b14}) by $\frac{dM(r)}{r}$ and
integrating from 0 to $R$, we finally obtain the generalized
virial theorem in warped DGP scenario as
\begin{eqnarray}
\left(1+\frac{\lambda_b}{6}\kappa_{5}^4\mu^2\right)2K+W+\frac{1}{3}\Lambda_4
I+{\cal W} = 0,\label{virial}
\end{eqnarray}
where
\begin{eqnarray}
W = -\frac{\kappa_4^2}{8\pi}\int^R_0\frac{M(r)}{r}
dM(r),\label{b17}
\end{eqnarray}
\begin{eqnarray}
{\cal W} = -\frac{\kappa_4^2}{2}\int^R_0{\cal M}(r)\rho r
dr,\label{b18}
\end{eqnarray}
and
\begin{eqnarray}
I = \int^R_0r^2dM(r),\label{b19}
\end{eqnarray}
where $W$ is the gravitational potential energy of the system. At
this point it is worth nothing that for $\mu=0$, we have $M_{DGP}
= 0$ and the virial theorem in the warped DGP brane-world is
reduced to the virial theorem in the RS brane scenario
\cite{Harko}
\begin{eqnarray}
2K+W+\frac{1}{3}\Lambda_4 I+W_{RS} = 0,\label{b16}
\end{eqnarray}
where
\begin{eqnarray}
{W}_{RS} = -\frac{\kappa_4^2}{2}\int^R_0{M}_{RS}(r)\rho r
dr.\label{b17}
\end{eqnarray}
As one can see the gravitational energy modified by $W_{RS}$ which
has its origin in the global bulk effect due to the ${\cal
E}_{\mu\nu}$ term. We can also recover the virial theorm in the
standard general relativity from equation (\ref{b16}) in the limit
$\lambda_b^{-1}\rightarrow0$. A alternative possibility in
recovering the four-dimensional virial theorem is to take the
limit $\kappa_5\rightarrow0$, while keeping the Newtonian
gravitational constant $\kappa_4^2$ finite [11].

In the case $\lambda_b = \Lambda^{(5)} = 0$, the virial theorem in
warped DGP brane-world is reduced to the virial theorem in DGP
brane scenario \cite{Shahidi}
\begin{eqnarray}
2K+W+W_{DGP} = 0,\label{b18}
\end{eqnarray}
where
\begin{eqnarray}
{W}_{DGP} = -\frac{\kappa_4^2}{2}\int^R_0{M}_{DGP}(r)\rho r
dr.\label{b19}
\end{eqnarray}
Note for a Minkowski DGP bulk space we have ${\cal E}_{\mu\nu} =
0$, thus in above equation $W_{RS} = 0$. There is difference
between our model with references \cite{Harko} and \cite{Shahidi}.
The virial theorem in warped DGP brane-world modified by both
$W_{RS}$ and $W_{DGP}$, which the first is due to the global bulk
effect whereas the second term has its origins in the induced
gravity on the brane due to quantum correction.

Now, we introduce the radii $R_{V}$ , $R_I$ and ${\cal R}$ as
\begin{eqnarray}
R_V = \frac{M^2(r)}{\int_0^R \frac{M(r)}{r}dM(r)},\label{b20}
\end{eqnarray}
\begin{eqnarray}
R_I^2 = \frac{\int_0^R r^2 dM(r)}{M(r)},\label{b21}
\end{eqnarray}
\begin{eqnarray}
{\cal R} = -\frac{\kappa_4^2}{8\pi}\frac{{\cal M}^2(r)} {{\cal
W}},\label{b22}
\end{eqnarray}
where $R_{V}$ is the virial radius and ${\cal R}$ is defined as
the geometrical radius of the clusters of galaxies. Defining the
virial mass as \cite{Jackson}
\begin{eqnarray}
2K = -\frac{\kappa_4^2}{8\pi}\frac{M_V^2}{R_V},\label{b23}
\end{eqnarray}
and using the following relations
\begin{eqnarray}
W = -\frac{\kappa_4^2}{8\pi}\frac{M^2}{R_V},\hspace{.5 cm}I =
MR_I^2,\label{b24}
\end{eqnarray}
the generalized virial theorem (\ref{virial}) is simplified as
\begin{eqnarray}
\left(1+\frac{\lambda_b}{6}\kappa_{5}^4\mu^2\right)\left(\frac{M_V}{M}\right)^2
= 1-\frac{8\pi\Lambda_4}{3\kappa_4^2}\frac{
R_VR_I^2}{M}+\left(\frac{\cal M}{M}\right)^2\left(\frac{R_V}{\cal
R}\right).\label{b25}
\end{eqnarray}
We have three types of mass in equation (\ref{b25}), namely, the
total baryonic mass of the system represented by $M$ (including
the baryonic mass of the intra-cluster gas and of the stars, other
particles like massive neutrinos), the virial mass represented by
$M_V$ and finally, the geometrical mass represented by ${\cal M}$.

On large distance scales associated with galaxies, we can ignore
the contribution of the effective cosmological constant to the
mass energy of the galaxy. Also, it is found that $M_{V}$ is
considerably greater than $M$ for most of the clusters and we can
neglect the term unitary in equation (\ref{b25}). Therefore, the
virial mass is given by
\begin{eqnarray}
M_V(r) \simeq {\cal M}(r)\left(\frac{R_{V}}{{\cal
R}}\right)^{1/2}.\label{b27}
\end{eqnarray}
This equation shows that the virial mass is proportional to the
geometrical mass.

\section{Astrophysical applications}

In this section, we emphasize that the astrophysical observations
together with the cosmological simulations have shown that the
virilized part of cluster is a measure of a fixed density such as
a critical density, $\rho_c(z)$ at a special red-shift, so that
$\rho_{V} = 3M_{V}/4\pi R_{V}^3 = \delta\rho_c(z)$, where $M_V$
and $R_V$ are the virial mass and radius, respectively. As is well
known, $\rho_c(z) = h^2(z)3H_0^2 /8\pi G$, where the Hubble
parameter is normalized to its local value, i.e., $h^2(z) =
\Omega_m(1+z)^3+ \Omega_\Lambda$, where $\Omega_m$ and
$\Omega_\Lambda$ are the mass density and dark energy density
parameters, respectively \cite{X}. By knowing the integrated mass
of the galaxy cluster as a function of the radius, one can
estimate the appropriate physical radius for the mass measurement.
The radii commonly used are $r_{200}$ or $r_{500}$. These radii
lie within the radii corresponding to the mean gravitational mass
density of the matter $\rho_{tot} = 200   \rho_{c}$ or $500
  \rho_c$. A useful radius is $r_{200}$ to find the virial mass. The
numerical values of the radius $r_{200}$ for the cluster NGC 4636
are in the ranges $r_{200} = 0.85$ $Mpc$ and $r_{200} = 4.49$
$Mpc$ for the cluster A2163, so one can deduce that a typical
value for $r_{200}$ is 2 $Mpc$. The masses corresponding to
$r_{200}$ and $r_{500}$ are denoted by $M_{200}$ and $M_{500}$,
respectively, and it is usually assumed that $M_V = M_{200}$ and
$R_V = r_{200}$ \cite{Y}.

\subsection{Geometrical mass estimated using the Jean's relation}

Now, we are going to obtain ${\cal M}(r)$ as a function of $r$ by
comparing the virial theorem results with the observational data
for galaxy cluster which can be obtained from the X-ray
observation of the gas in the cluster. The most of the baryonic
mass in clusters is in the gas form, therefore we assume that the
energy-density and pressure in $T_{\mu\nu}$ is that of a gas as
\begin{equation}\label{c1}
\rho = \rho_g(r),\hspace{.5 cm}p=p_g(r).
\end{equation}
In majority of clusters most of the baryonic mass is in the form
of the intra-cluster gas. The gas mass density $\rho_g(r)$
distribution can be fitted with the observational data by using
the following expression for the radial baryonic mass distribution
\cite{Y}
\begin{equation}\label{c2}
\rho_g(r) =
\rho_0\left(1+\frac{r^2}{r_c^2}\right)^{\frac{-3\beta}{2}},
\end{equation}
where $r_c$ is the core radius, and $\rho_0$ and $\beta$ are
cluster independent constants. A static spherically symmetric
system of collision-less particles that is in equilibrium, can be
described by the Jean's equation \cite{book}
\begin{equation}\label{c3}
\frac{d}{dr}\left[\rho_g(r)\sigma_r^2\right]+\frac{2\rho_g(r)}{r}
\left(\sigma_r^2-\sigma^2_{\theta,\varphi}\right) =
-\rho_g(r)\frac{d\Phi(r)}{dr},
\end{equation}
where $\Phi(r)$ is the gravitational potential, $\sigma_r$ and
$\sigma^2_{\theta,\varphi}$ are the mass-weighted velocity
dispersions in the radial and tangential directions. We assume
that the gas is isotropically distributed inside the cluster, so
$\sigma_r = \sigma_{\theta,\varphi}$. The gas pressure is related
to the velocity dispersion and gas profile density by $p_g =
\rho_g \sigma_r^2$ . By assumption that the gravitational field is
weak so that it satisfies the usual Poisson equation
$2\nabla^2\Phi \approx \kappa_4^2\rho_{tot}$, where $\rho_{tot}$
is the energy density including $\rho_g$ and other forms of
matter, like luminous matter and the geometrical matter, etc., the
Jean's equation becomes
\begin{equation}\label{c4}
\frac{dp_g(r)}{dr} = -\rho_g(r)\frac{d\Phi(r)}{dr} =
-\frac{\kappa_4^2 M_{tot}}{8\pi r^2}\rho_g(r),
\end{equation}
where $M_{tot}(r)$ is the total mass inside the radius $r$. The
observed X-ray emission from the hot ionized intra-cluster gas is
usually interpreted by assuming that the gas is in isothermal
equilibrium. Therefore we assume that the gas is in equilibrium
state having the equation of state $p_g(r) = \frac{k_B T_g}{\mu
m_p}\rho_g(r)$, where $k_B$ is Boltzmann constant, $T_g$ is the
gas temperature, $\mu = 0.61$ is the mean atomic weight of the
particles in the gas cluster and $m_p$ is the proton mass.
Equation (\ref{c4}) then gives
\begin{equation}\label{c5}
M_{tot}(r) = -\frac{8\pi k_B T_g}{\mu m_p \kappa_4^2} r^2
\frac{d}{dr}\ln\rho_g(r).
\end{equation}
Now, use of the density profile of the gas given by equation
(\ref{c2}) leads to the mass profile inside the cluster as
\begin{equation}\label{c6}
M_{tot}(r) = \frac{24\pi k_BT_g\beta} {\mu m_p\kappa_4^2 }
\frac{r^3} {r^2 + r_c^2}.
\end{equation}
On the other hand, using equation (\ref{b15}) and (\ref{M3}) we
can obtain another expression for the total mass
\begin{equation}\label{c7}
\frac{dM_{tot}(r)}{dr} = 4\pi r^2\rho_g(r) +
\frac{48\pi}{\kappa_4^2\lambda_b}U(r)r^2+\frac{8\pi\kappa_5^4\mu^2}{\kappa_4^2}\left[{\cal
P}(r)+3{\cal U}(r)\right]r^2,
\end{equation}
substituting equations (\ref{c6}) and (\ref{c2}) into equation
(\ref{c7}) we obtain the following expression
\begin{equation}\label{c8}
\frac{12}{\kappa_4^2\lambda_b}U(r)+\frac{2\kappa_5^2\mu^2}{\kappa_4^4}\left[{\cal
P}(r)+3{\cal U}(r)\right] = \frac{6k_BT_g\beta} {\mu m_p
\kappa_4^2} \frac{r^2+3r_c^2}{
(r^2+r_c^2)^2}-\rho_0\left(1+\frac{r^2}{r_c^2}\right)^{-3\beta/2}.
\end{equation}
Finally, substituting above equation into equation (\ref{M3}), in
the limit $r>>r_c$ considered here, we obtain the following
geometrical mass
\begin{equation}\label{c9}
{\cal M}(r) \simeq \left[\frac{24\pi k_BT_g\beta} {\mu m_p
\kappa_4^2}-4\pi\rho_0r_c^{3\beta}\frac{r^{2-3\beta}}{3(1-\beta)}\right]
r,
\end{equation}
which includes both the local and non-local bulk effects.
Observations show that the intra cluster gas has a small
contribution to the total mass \cite{X, Y, W, Z}, thus we can
neglect the contribution of the gas to the geometrical mass and
rewrite equations (\ref{c9}) as
\begin{equation}\label{c}
{\cal M}(r) \simeq \left(\frac{24\pi k_BT_g\beta} {\mu m_p
\kappa_4^2}\right) r.
\end{equation}
Now Let us estimate the value of ${\cal M}(r)$. First we note that
$k_B T_g \approx 5 KeV$ for most clusters. The virial radius of
the clusters of galaxies is usually assumed to be $r_{200}$,
indicating the radius for which the energy density of the cluster
becomes $\rho_{200} = 200 \rho_{cr}$, where $\rho_{cr} =
4.6975\times10^{-30}h^2_{50} gr/cm^{3}$ \cite{Y}. Using equation
(\ref{c9}) we find
\begin{equation}\label{c10}
r_{cr} = 91.33 {\beta}^{1/2}\left(\frac{k_BT_g}{5
KeV}\right)^{\frac{1}{2}} h^{-1}_{50} Mpc.
\end{equation}
The total geometrical mass corresponding to this value is
\begin{equation}\label{c12}
{\cal M}{(r)} =
4.83\times10^{16}{\beta}^{3/2}\left(\frac{k_BT_g}{5
KeV}\right)^{\frac{1}{2}} h^{-1}_{50} M_{\odot},
\end{equation}
which is consistent with the observational values for the virial
mass of clusters \cite{Y}.
\subsection{Radial velocity dispersion in galactic clusters}

Radial velocity dispersion in galactic clusters plays an important
role in estimating the virial mass of the clusters. It can be
expressed in terms of the virial mass as \cite{Z}
\begin{equation}\label{d0}
M_V = \frac{3}{G}\sigma_1^2 R_V.
\end{equation}
Assuming that the velocity distribution in the cluster is
isotropic, we have $\langle{v^2}\rangle = \langle{v_r^2}\rangle +
\langle{v_\theta^2}\rangle + \langle{v_\varphi^2}\rangle =
3\langle{v_r^2}\rangle = 3\sigma_r^2$, the radial velocity
dispersion $\sigma_r^2$ for clusters in warped DGP model can be
obtained from equation (\ref{b6}) as
\begin{equation}\label{d1}
\frac{d} {dr} (\rho\sigma_r^2 ) + \frac{1} {2} \rho\lambda' = 0,
\end{equation}
where $\sigma_1$ and $\sigma_r$ are related by $3\sigma^2_1 =
\sigma^2_r$. On the other hand, by neglecting the cosmological
constant the Einstein field equation (\ref{b13}) becomes
\begin{eqnarray}
\left(1+\frac{\lambda_b}{6}\kappa_{5}^4\mu^2\right)\frac{1}{2r^2}\frac{\partial}{\partial
r}\left(r^2\frac{\partial \lambda}{\partial r}\right) =
\frac{\kappa_4^2}{2}\rho+{\kappa_5^4\mu^2}\left[{\cal P}(r)+3{\cal
U}(r)\right]+\frac{6}{\kappa_4^2\lambda_b}U(r).\label{d3}
\end{eqnarray}
Integrating, we obtain
\begin{eqnarray}
\left(1+\frac{\lambda_b}{6}\kappa_{5}^4\mu^2\right)\frac{1}{2}\left(r^2\frac{\partial
\lambda}{\partial r}\right) = \frac{\kappa_4^2}{8\pi
}M(r)+\frac{\kappa_4^2}{8\pi }{\cal M}(r)+{{\cal C}_1},\label{d4}
\end{eqnarray}
where ${\cal C}_1$ is an integration constant. By eliminating
$\lambda'$ from equations (\ref{d1}) and (\ref{d4}), we obtain
\begin{equation}\label{d5}
\left(1+\frac{\lambda_b}{6}\kappa_{5}^4\mu^2\right)\frac{d}{dr}(\rho\sigma_r^2)
= -\frac{\kappa_4^2M(r)}{8\pi r^2}\rho(r)-\frac{\kappa_4^2{\cal
M}(r)}{8\pi r^2}\rho(r)-\frac{{\cal C}_1}{r^2}\rho(r).
\end{equation}
Integration now gives the following solution
\begin{equation}\label{d6}
\left(1+\frac{\lambda_b}{6}\kappa_{5}^4\mu^2\right)\sigma_r^2 =
-\frac{1}{\rho}\int\frac{\kappa_4^2M(r)}{8\pi
r^2}\rho(r)dr-\frac{1}{\rho}\int\frac{\kappa_4^2{\cal M}(r)}{8\pi
r^2}\rho(r)dr-\frac{1}{\rho}\int\frac{{\cal
C}_1}{r^2}\rho(r)dr-\frac{{\cal C}_2}{\rho},
\end{equation}
where ${\cal C}_2$ is an integration constant. For most clusters
$\beta\geq\frac{2}{3}$ and therefore, in the limit $r>> r_c$, the
gas density profile (\ref{c2}) can be written as \cite{Harko}
\begin{equation}\label{d7}
\rho_g(r) = \rho_0 \left(\frac{r}{r_c}\right)^{-3\beta},\hspace
{0.5 cm}\beta\geq\frac{2}{3}.
\end{equation}
Now, substituting equations (\ref{b15}), (\ref{c}) and equation
(\ref{d7}) into equation (\ref{d6}), for $\beta\neq 1$ we obtain
\begin{equation}\label{d9}
\left(1+\frac{\lambda_b}{6}\kappa_{5}^4\mu^2\right)\sigma_r^2 =
-\frac{\rho_0\kappa_4^2}{12(1-\beta)(1-3\beta)}r^2\left(\frac{r}{r_c}\right)^{-3\beta}+\frac{k_BT_g}{\mu
m_p}+\frac{{\cal C}_1}{(1+3\beta)}\frac{1}{r}-\frac{{\cal
C}_2}{\rho_0}\left(\frac{r}{r_c}\right)^{3\beta},
\end{equation}
and for $\beta = 1$ we find
\begin{equation}\label{d9}
\left(1+\frac{\lambda_b}{6}\kappa_{5}^4\mu^2\right)\sigma_r^2 =
\frac{\rho_0 \kappa_4^2  }{8}r_c^3\left(\frac{1}{4r^4}+\frac{\ln
r}{r^4}\right)\left(\frac{r}{r_c}\right)^{3}+\frac{k_BT_g}{\mu
m_p}+\frac{{\cal C}_1}{4}\frac{1}{r}-\frac{{\cal
C}_2}{\rho_0}\left(\frac{r}{r_c}\right)^{3}.
\end{equation}
As is well-known, the simple form $\sigma_r^2 (r) = B/(r+b)$ for
the radial velocity dispersion and the relation $\rho (r) = A/r (r
+ a)^2$ for the density of the galaxies in cluster, with $B$, $b$,
$a$ and $A$ constants, can be used to fit the observational data
\cite{Z}. For $r<<a$, $\rho(r)\simeq A/{r}$, while for $r>>a$,
$\rho(r)$ behaves like $\rho(r)\simeq A/{r^3}$. Here, our
expression for $\sigma_r^2$ can be also used to fit the
observational data. Therefore, the comparison of the observed
velocity dispersion profiles of the galaxy clusters and the
velocity dispersion profiles predicted by the warped DGP
brane-world model may give a powerful method to discriminate
between the different theoretical scenarios.
\begin{figure}
\centerline{\begin{tabular}{ccc}
\epsfig{figure=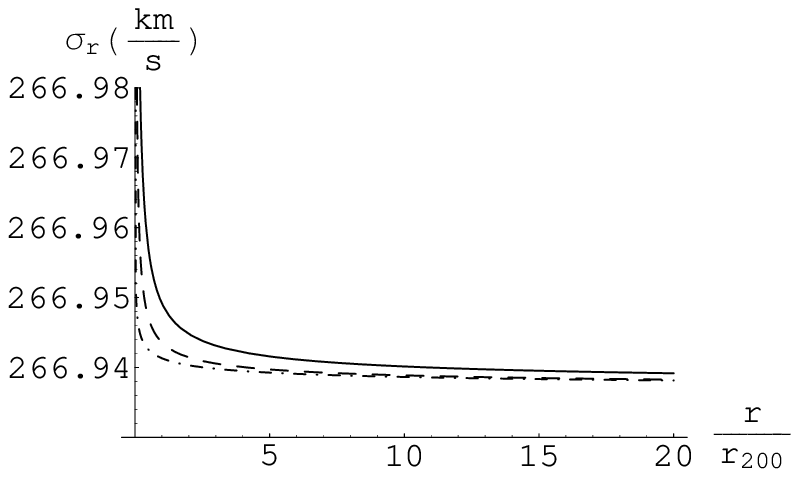,width=7cm}\hspace{20mm}\epsfig{figure=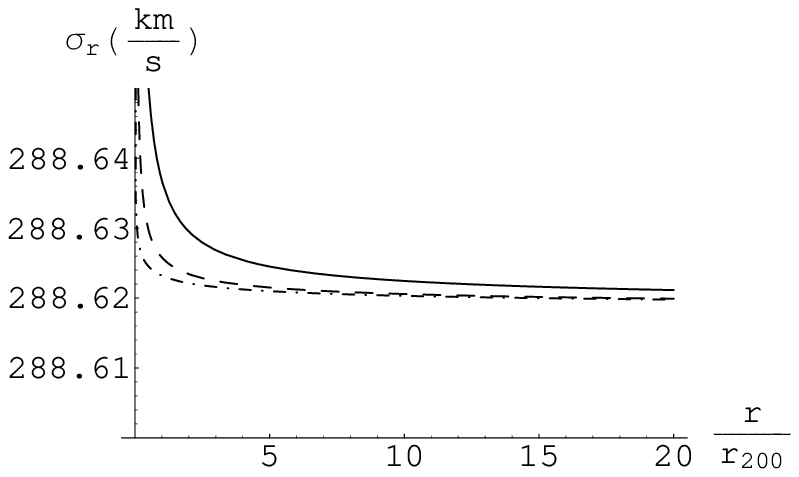,width=7cm}
\end{tabular} } \caption{\footnotesize  Left, The radial velocity
dispersion in warped DGP brane-world model and right, the same
parameter in DGP brane-world scenario for the NGC 5813 cluster
with $\beta = 0.766$, $r_{200} = 0.87Mpc$, $r_c = 25Kpc$, $k_BT_g
= 0.52KeV$ and ${\cal C}_1 = 0.503, 1.005\times 10^{8},
2.011\times 10^{8}\hspace{0.1cm}M_{\odot}$, ${\cal C}_2 = 0.02,
0.03, 0.04\hspace{0.1cm}{M_{\odot}^2}/{Kpc}^{4}$ for solid, dashed
and dot-dashed curves, respectively.}\label{fig1}
\end{figure}

Finally, we compare the radial velocity dispersion in warped DGP
brane-world model with the radial velocity dispersion in the other
theoretical models. As we noted before for $\Lambda^{(5)} =
\lambda_b = 0$, the warped DGP model reduce to the DGP model and
equation (\ref{d9}) for $\beta=1$ reduce to
\begin{equation}\label{n1}
\sigma_r^2 = \frac{\rho_0 \kappa_4^2
}{8}r_c^3\left(\frac{1}{4r^4}+\frac{\ln
r}{r^4}\right)\left(\frac{r}{r_c}\right)^{3}+\frac{k_BT_g}{\mu
m_p}+\frac{{\cal C}_1}{4}\frac{1}{r}-\frac{{\cal
C}_2}{\rho_0}\left(\frac{r}{r_c}\right)^{3},
\end{equation}
which is the radial velocity dispersion in DGP model, equation
(65) in \cite{Shahidi}. The radial velocity dispersion of galaxy
clusters in Palatini $f(R)$ gravity for $\gamma = 3$, which is
corresponding to $\beta = 1$, is also presented as \cite{Sefied}
\begin{equation}\label{n2}
\sigma_r^2 = -r^{3}\int\frac{F'}{2F} r^{-3} dr+\frac{k_BT_g}{\mu
m_p}+{\pi G \rho_0}\left(\frac{1}{4r}+{\frac{\ln
r}{r}}\right)+\frac{c}{4}\frac{1}{r}-\frac{c'}{\rho_0}r^3,
\end{equation}
where $F(R) = \frac{df(R)}{dR}$. For $f(R) = R$ this relation
reduces to
\begin{equation}\label{n3} \sigma_r^2 =
\frac{\kappa_4^2\rho_0}{8}\left(\frac{1}{4r}+{\frac{\ln
r}{r}}\right)+\frac{k_BT_g}{\mu
m_p}+\frac{c}{4}\frac{1}{r}-\frac{c'}{\rho_0}r^3,
\end{equation}
which is equation (\ref{d9}) with $8\pi G = \kappa_4^2$ and
$\Lambda^{(5)} = \lambda_b = 0$. The same relation has been also
obtained in Randall-Sundrum II model with this difference that the
origin of geometrical mass in $f(R)$ gravity is the extra terms in
the Einstein-Hilbert action whereas in the latter is the global
bulk effect. In figure 2 we have plotted the radial velocity
dispersion for the cluster NGC 5813. The numerical values of it
are in the ranges $\beta = 0.766$, $r_c = 25 Kpc$, $k_BT_g = 0.52
Kev$, $r_{200} = 0.087 Mpc$ \cite{Y} and the radial velocity is
about $240 km/s$ \cite{An}. As one can see the radial velocity
dispersion in warped DGP brane-world is compatible with the
observed profiles and for the same value of constants ${\cal C}_1$
and ${\cal C}_2$ is slower than the DGP brane-world model.

\section{Conclusions}
The virial theorem plays an important role in astrophysics because
of its generality and wide range of applications. One of the
important results which can be obtained with the use of the virial
theorem is to derive the mean density of astrophysical objects
such as galaxy clusters and it can be used to predict the total
mass of the clusters of galaxies. In the present paper, using the
collision-less Boltzmann equation, we have obtained the
generalized virial theorem within the context of the warped DGP
brane-world model. The additional geometric terms due to the
induced curvature term on the brane and non-local bulk effect in
the modified gravitational field equations provide an effective
contribution to the gravitational energy, equation (\ref{b14}),
which may be used to explain the well-known virial theorem mass
discrepancy in clusters of galaxies. Finally, we have compared the
virial theorem results with the observational data for galaxy
cluster which can be obtained from the X-ray observation of the
gas in the cluster and expressed the geometrical mass in term of
observational quantities, like the temperature and the gas profile
density.

\section*{Acknowledgment}

We would like to thank the research office of Qom University for
financial supports. We are also indebted to two anonymous referees
of the Physical Review D for valuable comments.

\end{document}